\documentclass[reprint,amsmath,amssymb,aps]{revtex4-2}

\usepackage{graphicx}
\usepackage{dcolumn}
\usepackage{bm}
\usepackage{color}

\usepackage{calc}
\newsavebox\CBox
\newcommand\hcancel[2][0.5pt]{%
  \ifmmode\sbox\CBox{$#2$}\else\sbox\CBox{#2}\fi%
  \makebox[0pt][l]{\usebox\CBox}%
  \raisebox{-.35mm}{\rotatebox{25}{\rule[0.5\ht\CBox-#1/2]{\wd\CBox}{#1}}
  }}

\def\LS {\ensuremath{\Lambda\bar\Sigma^0+{\rm c.c.}}}
\def\be {\begin{eqnarray}}
\def\en {\end{eqnarray}}
\def\ne {\nonumber\end{eqnarray}}
\def\no {\nonumber}
\def\ee {\ensuremath{e^+e^-}}
\def\bb {\ensuremath{B_1\bar B_2+{\rm c.c.}}}
\def\lt {\ensuremath{\left(}}
\def\rt {\ensuremath{\right)}}

\def\lab {\ensuremath{\left|}}
\def\rab {\ensuremath{\right|}}
\def\lgr {\ensuremath{\left\{}}

\def\ds {\displaystyle}
\def\noi {\ensuremath{\scalebox{.8}{\hcancel{$I$}}}}

\def\BB {\ensuremath{B_1B_2}}
\def\lsz {\ensuremath{\Lambda\Sigma^0}}
\def\jp {\ensuremath{J/\psi}}
\def\pp {\ensuremath{\psi(2S)}}

\begin{document}

\preprint{APS/123-QED}

\title{Isospin strikes back}
\thanks{Based on Ref.~\cite{Ferroli:2020xnv}.}%

\author{Francesco Rosini}
 \email{francesco.rosini@pg.infn.it}
\author{Simone Pacetti}%
 \email{simone.pacetti@unipg.it}
\affiliation{%
Dipartimento di Fisica e Geologia, INFN Sezione di Perugia, 06123 Perugia, Italy
}%

\begin{abstract}
Assuming isospin conservation, the decay of a $c\bar c$ vector meson into the $\LS$ final state is purely electromagnetic. At the leading order, the $c\bar c$ vector meson first converts into a virtual photon that, then produces the \LS\ final state. Moreover, such a mechanism, i.e., the virtual photon coupling to \LS, is the solely intermediate process through which, in Born approximation, the reaction $e^+e^-\to\LS$ does proceed. It follows that any significant difference between the amplitudes of the processes $c\bar c\to\LS$ and $e^+e^-\to\LS$ at the $c\bar c$ mass must be ascribed to an isospin-violating contribution in the $c\bar c$ decay. 
\\
In Ref.~\cite{Ferroli:2020xnv} we studied the decay of the $\psi(2S)$ vector meson into \LS\ and, on the light of the large branching fraction
\be
{\rm BR}_{18}(\psi(2S)\to\LS)=(1.23\pm0.24)\times 10^{-5}\,,
\ne
published in the 2018 edition of the Review of Particle Physics~\cite{PhysRevD.98.030001}, we claimed either the presence of a significant isospin-violating contribution or, with a lesser emphasis, a "not complete reliability of the only available datum". In any case, we propose a new measurement. Apparently, our second and considered less serious hypothesis was the right one, indeed the  branching fraction published in the 2024 edition of the Review of Particle Physics~\cite{PhysRevD.110.030001} is
\be
{\rm BR}(\psi(2S)\to\LS)=(1.6\pm0.7)\times 10^{-6}\,,
\ne
more than seven times lower with the error that increased from $\sim 20\%$ to $\sim 45\%$.
\end{abstract}

\maketitle
\section{Parameterizations of the amplitudes
\label{sec:parameterizations}}
\begin{figure}[h]
	\begin{center}
		\includegraphics[width = .8\columnwidth]{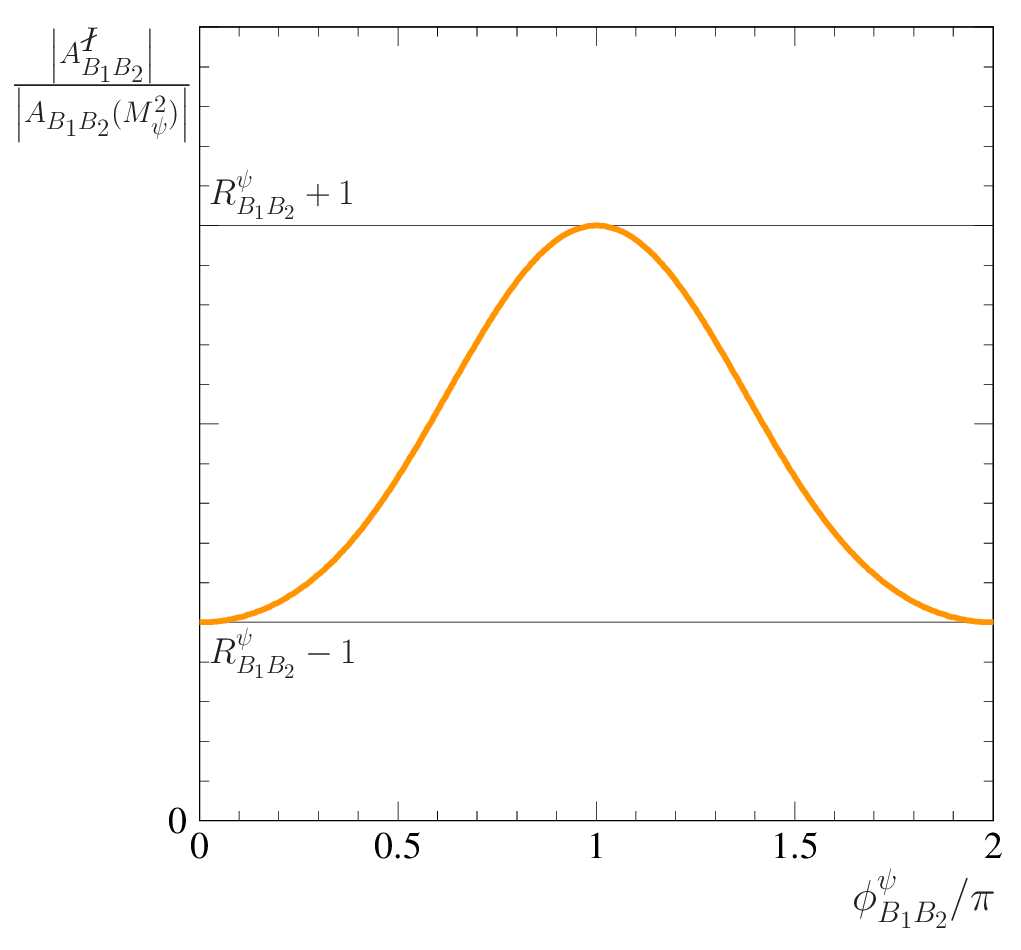}
		\label{fig:example}
		\caption{Relative isospin-violating-to-pure-electromagnetic intensity in arbitrary units as a function of the relative phase in units of $\pi$ radians.}
	\end{center}
\end{figure}

The processes under consideration are the decay of a $c\bar c$ vector meson, called $\psi$, into a baryon-antibaryon \bb\ final state, and the \ee\ annihilation which produces the same final state. The reactions are
\be
\psi\to\bb\,, \ \ \ \ee\to\bb\,.
\ne
In general, at the leading-order, the $\psi$ and $\gamma$-hyperons currents are
\be
&&\displaystyle
J_\mu^\psi=\bar u(k_1)G_\mu[g_E^{\BB},g_M^{\BB}]v(k_2)\,,
\no\\
&&\displaystyle
J_\mu^\gamma=\bar u(k_1)G_\mu[G^{\BB}_E(q^2),G^{\BB}_M(q^2)]v(k_2)\,,
\ne
where $u$ and $v$ are the Dirac spinors,	 $k_1$ and $k_2$ are the four-momenta of the hyperons, $q=k_1+k_2$, $g_E^{\BB}$ and $g_M^{\BB}$ are the so-called psionic coupling constants, while $G^{\BB}_E(q^2)$ and $G^{\BB}_M(q^2)$ are Sachs form factors. The Lorentz four-vector $G_\mu$ in both the currents has the same expression, in that, the second can be obtained from the first by the substitutions: $g_{E,M}^{\BB}\to G^{\BB}_{E,M}(q^2)$ and vice versa. 
\\
The rate of the decay $\psi\to\bb$ and the cross section of the annihilation $\ee\to\bb$ are
\be
\Gamma^\psi_{B_1B_2}&=&
\beta_{B_1B_2}(M_\psi^2)\Gamma^\psi_{\mu\mu}\no\\
&&\cdot\lt\frac{2M_{B_1B_2}^2}{M_\psi^2}\lab g_E^{\BB}\rab^2+\lab g_M^{\BB}\rab^2\rt\,,
\label{eq:FF}
\\
\sigma_{B_1B_2}(q^2)&=&\frac{4\pi\alpha^2\beta_{B_1B_2}(q^2)}{3q^2}\no\\
&&\cdot\lt\frac{2M_{B_1B_2}^2}{q^2}\lab G^{\BB}_E(q^2)\rab^2+\lab G^{\BB}_M(q^2)\rab^2\rt\,,
\ne
where $\beta_{B_1B_2}(q^2)=\sqrt{1-{4M_{B_1B_2}^2}/{q^2}}$ is the velocity of both out-going baryons in the baryon-anti-baryon center-of-mass frame, $M_\psi$ is the mass of the vector meson $\psi$, $\Gamma^\psi_{\mu\mu}$ is the decay rate for $\psi\to\mu^+\mu^-$, which accounts for the electromagnetic coupling of the vector meson $\psi$ with the virtual photon, $\alpha$ is the fine structure constant, and $M_{B_1B_2}=\sqrt{ {\lt M_{B_1}^2+M_{B_2}^2\rt}/{2}-{\lt M_{B_1}^2-M_{B_2}^2\rt^2}/(4q^2)}$, which, in the case of $B_1=B_2\equiv B$, is the baryon mass, i.e., $M_{B_1B_2}=M_B$. 
\\
To make a straightforward comparison we define the effective unique psionic coupling constant $A_{\BB}^\psi$ and the effective form factor $A_{\BB}(q^2)$ as
\be
&&\ds |A_{\BB}^\psi|=
\sqrt{\frac{2M_{B_1B_2}^2}{M_\psi^2}\lab g_E^{\BB}\rab^2
+\lab g_M^{\BB}\rab^2}\,,
\label{eq:eff}\\
&&\ds
|A_{\BB}(q^2)|=\sqrt{\frac{2M_{B_1B_2}^2}{q^2}\lab G_E^{\BB}(q^2)\rab^2
+\lab G_M^{\BB}(q^2)\rab^2}\,,
\ne
in terms of which the decay rate and cross section read
\be
&&\ds
\Gamma^\psi_{B_1B_2}=\beta_{B_1B_2}(M_\psi^2)\Gamma^\psi_{\mu\mu}\lab A_{\BB}^\psi\rab^2
\,,\label{eq:rate-xs}\\
&&\ds
\sigma_{B_1B_2}(q^2)=\frac{4\pi\alpha^2\beta_{B_1B_2}(q^2)}{3q^2}\lab A_{\BB}(q^2)\rab^2\,.
\ne
In the case of a pure isovector final state, as for $(B_1,B_2)=(\Lambda,\Sigma^0)$, and assuming isospin conservation, the psionic coupling constants coincide with the Sachs form factors at the vector meson mass, i.e., 
\be
g_{E,M}^{\BB}=G_{E,M}^{\BB}(M_\psi^2)\,.
\label{eq:conserva}
\en
An isospin-violating contribution can be parametrized by the pair of coupling constants $g_{E,\scalebox{.8}{\hcancel{$I$}}}$ and $g_{M,\scalebox{.8}{\hcancel{$I$}}}$ that sum up to the isospin-conserving ones to give the total psionic coupling constants as
\be
g_{E,M,\rm tot}^{{\BB}}=g_{E,M}^{\BB}+g_{E,M,\scalebox{.8}{\hcancel{$I$}}}^{\BB}\,.
\ne
In terms of the effective coupling constant and the effective form factor defined in Eq.~\eqref{eq:eff}, the total coupling constant is defined as 
\be
A^{\psi}_{{\BB},\rm tot}
=A_{\BB}^\psi+A_{\BB}^{\scalebox{.8}{\hcancel{$I$}}}\!\!
=A_{\BB}(M^2_\psi)+A_{\BB}^{\scalebox{.8}{\hcancel{$I$}}}\,,
\ne
where, as a consequence of Eq.~\eqref{eq:conserva}, the effective isospin-conserving psionic coupling constant coincides with the effective form factor at the $\psi$ mass.
\\
It follows that, by allowing the isospin violation, which can occur only in the decay process $\psi\to\LS$, since the reaction $\ee\to\LS$ is purely electromagnetic, the expression of the decay rate of Eq.~\eqref{eq:rate-xs} becomes
\be
\Gamma_{B_1B_2}^\psi
&=&
\beta_{B_1B_2}(M_\psi^2)\Gamma^\psi_{\mu\mu}\lab A^\psi_{\BB,\rm tot}\rab^2
\no\\
&=&
\beta_{B_1B_2}(M_\psi^2)\Gamma^\psi_{\mu\mu}
\lab A_{\BB}(M^2_\psi)+A_{\BB}^{\scalebox{.8}{\hcancel{$I$}}}\rab^2
\,.\label{eq:rate-i}
\en
The experimental value of the modulus of the ratio between the total effective coupling constant $A^\psi_{\BB,\rm tot}$ and the purely electromagnetic one $A_{\BB}^{\psi}=A_{\BB}(M_\psi^2)$ can be obtained from the measurements of the cross section at the $\psi$ mass and the decay rate. Using the cross section formula of Eq.~\eqref{eq:rate-xs} at $q^2=M_\psi^2$ and that for the decay rate of Eq.~\eqref{eq:rate-i}, the ratio reads
\be
R_{B_1B_2}^\psi &\equiv& {\lab A_{\BB}(M_\psi^2)+A_{\BB}^{\noi}\rab}/{\lab A_{\BB}(M_\psi^2)\rab}\no
\\&\equiv& \sqrt{\frac{4\pi\alpha^2}{3M_\psi^2\Gamma_{\mu\mu}^\psi}\frac{\Gamma^\psi_{B_1B_2}}{\sigma_{B_1B_2}(M_\psi^2)}}\,.
\label{eq:R}
\en
The intensity and hence the modulus of the isospin-violating coupling constant can be extracted from the experimental value $R_{B_1B_2}^\psi$ as a function of its relative phase to the electromagnetic coupling constant $A_{\BB}(M_\psi^2)$. Indeed, indicating with $\phi_{\BB}^\psi$ such a phase, we have
\be
\lab A_{\BB}(M_\psi^2)+A^{\noi}_{\BB}\rab^2
&=&
\lab A_{\BB}(M_\psi^2)\rab^2+\lab A_{\BB}^{\noi}\rab^2
\no\\
&&
\hspace{-20mm}+2\lab A_{\BB}(M_\psi^2)\rab\lab A_{\BB}^{\noi}\rab\cos\lt \phi_{\BB}^\psi\rt\,,
\ne
dividing by $\lab A_{\BB}(M_\psi^2)\rab^2$ and solving for the ratio $\lab A^{\noi}_{\BB}\rab/\lab A_{\BB}(M_\psi^2)\rab$, i.e., the relative isospin-violating-to-pure-electromagnetic intensity, we get
\be
{\lab A_{\BB}^{\noi}\rab}/{\lab A_{\BB}(M_\psi^2)\rab}&=&
\sqrt{\lt R_{B_1B_2}^\psi\rt^2-\sin^2\lt \phi_{\BB}^\psi\rt}
\no\\&&
-\cos\lt\phi_{\BB}^\psi\rt\,.
\label{eq:relativo}
\en
Figure~\ref{fig:example} shows the generic behavior of the ratio $\lab A_{\BB}^{\noi}\rab /\lab A_{\BB}(M_\psi^2)\rab$ as a function of the relative phase $\phi_\psi$. It is symmetric with respect to $\phi_{\BB}^\psi=\pi$, and it is minimum and equal to $R_{B_1B_2}^\psi-1$ at $\phi_{\BB}^\psi=0$ when the interference between the two coupling constants is  constructive and reaches the maximum value of $R_{B_1B_2}^\psi+1$ at $\phi_{\BB}^\psi=\pi$, when, instead, the interference is destructive.     
\section{The scaled cross section}
\label{sec:scaled-xs}
To extract the ratio $R_{B_1B_2}^\psi$ defined in Eq.~\eqref{eq:R}, the experimental values of the decay rate $\Gamma^\psi_{B_1B_2}$ and the cross section $\sigma_{B_1B_2}(q^2)$ are needed. While the formers are easily achievable in the Review of Particle Physics, e.g., Ref.~\cite{PhysRevD.110.030001} is the 2024 edition, for the cross-section values there are not so well-organized and precise sources of information. Mainly because cross sections even for a specific reaction are functions of energy and hence, they are not a single datum but a set of data having number of points and energy values characteristic of each experiment.
\\
To collect as much as possible information on the modulus of the effective form factor $A_{\BB}(q^2)$, for a specific final state \bb, we use cross section data of all the $\BB$ pairs of neutral baryons belonging to the spin-1/2 SU(3) baryon octet~\cite{BALDINIFERROLI2019135041}. This can be done by considering the SU(3)-symmetry breaking terms due to the electromagnetic effects of the leading-order Lagrangian density describing the decay $\psi\to\bb${\color{red}.} These terms define the effective electromagnetic coupling constants, i.e., the amplitudes for the reactions $\psi\to\gamma^*\to\bb$. In particular, the amplitudes of reactions producing pairs of neutral baryon-anti-baryon are proportional to the same coupling constant by known coefficients.
\\ 
As a consequence, the effective form factors of these neutral baryon-anti-baryon pairs and hence the cross sections can be expressed in terms of the effective form factor of an only pair scaled by the corresponding coefficient.
\\
Using the pair $(\Lambda,\Sigma^0)$ as the reference, the effective form factors of the other four neural baryon-anti-baryon pairs are
\be
A_{B_1B_2}(q^2)&=&N_{B_1B_2}A_{\Lambda\Sigma^0}(q^2)\,,
\label{eq:scaled0}\en
with $\lt N_{nn},N_{\Lambda\Lambda},N_{\Sigma^0\Sigma^0},N_{\Xi^0\Xi^0}\rt
=\lt-2,-1,1,-2\rt/\sqrt{3}$.
The scaled cross section is defined as~\cite{Ferroli:2020xnv}
\be
\tilde\sigma(q^2)=\frac{\sigma_{B_1B_2}(q^2)}{N_{B_1B_2}^2\beta_{B_1B_2}(q^2)}
=
\frac{4\pi\alpha^2}{3q^2}\lab A_{\Lambda\Sigma_0}(q^2)\rab^2\,,
\ne
and corresponds to the cross section of the annihilation $\ee\to\LS$ with the baryon velocity equal to one.
\\
The data sets used to extract the scaled cross section values are on all four neutral baryon-anti-baryon channels, besides that under study, namely: $(B_1,B_2)=(\Lambda,\Sigma^0)$. The measurements were performed by the following experiments:
\begin{itemize}
	\item BaBar on the reactions: $\ee\to\Lambda\bar\Lambda$, $\ee\to\Sigma^0\bar\Sigma^0$ and $\ee\to\LS$~\cite{BaBar:2007fsu};
	\item Belle on the reaction: $\ee\to\Sigma^0\bar\Sigma^0$~\cite{Belle:2022dvb};
	\item BESIII on the reactions: $\ee\to\LS$~\cite{BESIII:2023pfv}, $\ee\to\Lambda\bar\Lambda$~\cite{PhysRevLett.123.122003}, $\ee\to n\bar n$~\cite{BESIII:2021tbq}, and $\ee\to\Xi^0\bar\Xi^0$~\cite{BESIII:2024ues}.
\end{itemize}
\begin{figure}[h]
	\begin{center}
		\includegraphics[width = .8\columnwidth]{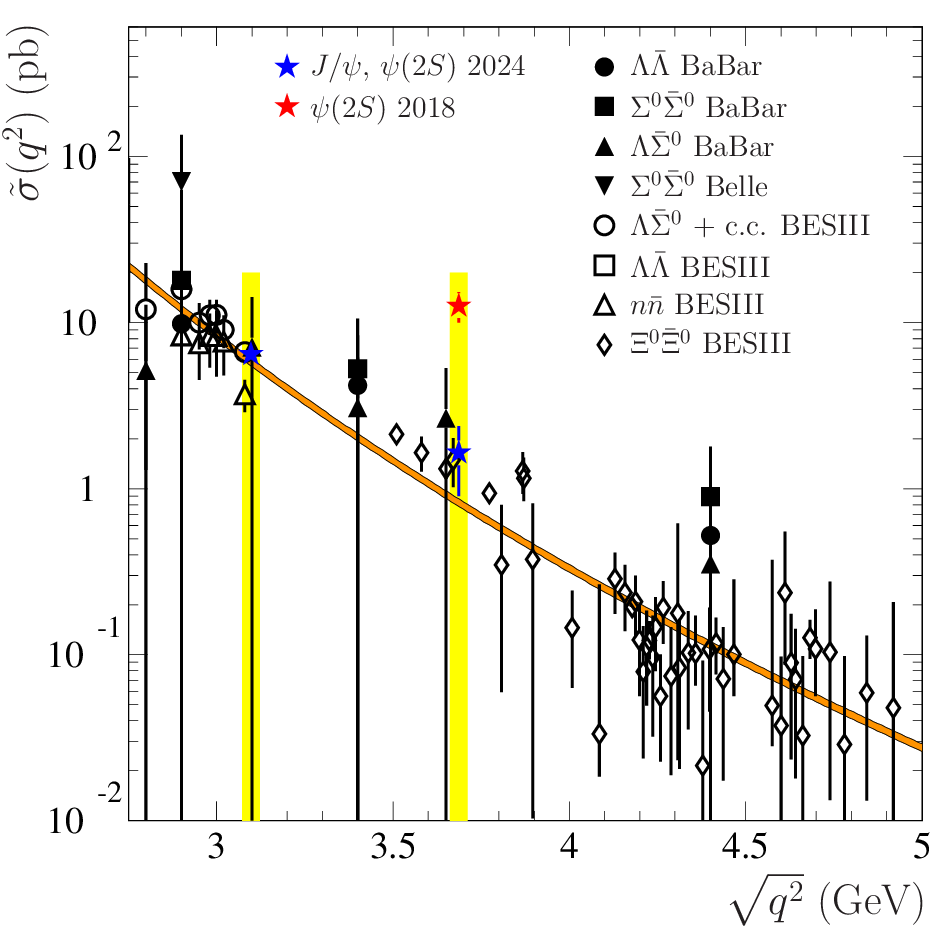}\vspace{-5mm}
		\label{fig:scaled-xs}
		\caption{Data on the scaled cross section for the reactions $\ee\to n\bar n$, $\Lambda\bar\Lambda$, $\Sigma^0\bar\Sigma^0$, $\Xi^0\bar\Xi^0$ and \LS, collected by the experiments: BaBar~\cite{BaBar:2007fsu}, Belle~\cite{Belle:2022dvb} and BESIII~\cite{BESIII:2023pfv,PhysRevLett.123.122003,BESIII:2021tbq,BESIII:2024ues}. The blue stars indicate the values of the scaled cross section at the $J/\psi$ and $\psi(2S)$ masses extracted by the the 2024~\cite{PhysRevD.110.030001} rates of their decays into \LS\ The red star is the scaled cross section an the $\psi(2S)$ mass obtained with the 2018~\cite{PhysRevD.98.030001} decay rate. The orange band represents the fit results including the errors. Energy regions 50 MeV-wide around the \jp\ and \pp\ masses are highlighted in yellow.}
	\end{center}
\end{figure}
Following the procedure defined in Ref.~\cite{Ferroli:2020xnv}, to extract the values of the modulus of the effective form factor $A_{\lsz}(q^2)$ at the $\jp$ and $\pp$ masses	, we assume the high-$q^2$-power law behavior predicted by the perturbative QCD~\cite{Matveev:1973uz,Brodsky:1973kr} and use the fitting function
\be
\tilde\sigma_{\rm fit}(q^2)=\frac{A}{(q^2)^5\lt\pi^2+\ln^2\lt q^2/\Lambda_{\rm QCD}^2\rt\rt^2}\,.
\label{eq:fit}
\en
This function, which accounts for the first order QCD-logarithmic correction, depends on only one free parameter $A$, the QCD scale is fixed at $\Lambda_{\rm QCD}=0.35$ GeV.  The best value of the parameter $A$ is obtained by minimizing the $\chi^2$
\be
\chi^2=\sum_{j=1}^N\lt\frac{\tilde\sigma_{\rm fit}(q^2_j)-\tilde\sigma_j}{\delta\tilde\sigma_j}\rt^2,\ \ 
\lgr\begin{array}{rcl}\!\!\!
\tilde\sigma_j\!\!\!\!&=&\!\!\!\!\ds\frac{\sigma_{B_1B_2,j}}{N_{B_1B_2}^2\beta_{B_1B_2}(q_j^2)}\vspace{2mm}\\
\delta\tilde\sigma_j\!\!\!&=&\!\!\!\ds\frac{\delta\sigma_{B_1B_2,j}}{N_{B_1B_2}^2\beta_{B_1B_2}(q_j^2)}
\\
\end{array}\right.,
\ne 
where $\sigma_{B_1B_2,j}\pm\delta\sigma_{B_1B_2,j}$ is the experimental value of the cross section of the annihilation $\ee\to\bb$ measured at $q^2=q_j^2$, for all data sets with $(B_1,B_2)=(n,n)$, $(\Lambda,\Lambda)$, $(\Sigma^0,\Sigma^0)$, $(\Xi^0,\Xi^0)$, $(\Lambda,\Sigma^0)$, and with $q_j^2\ge (2.8 \,{\rm GeV})^2$. This cut-off for the four-momentum transfer squared has been adopted to exclude the production threshold region where the behavior of the cross section does not follow the power-law predicted by perturbative QCD. The total number of data points considered for the minimization is $D=70$ and the obtained best value for the parameter $A$ is
	\be
	A=(3.92\pm 0.21 )\times 10^8\,\,{\rm GeV^{10}\,pb}\,.
	\label{eq:A-best}
	\en
The black symbols of Fig.~\ref{fig:scaled-xs} represent the data of all the processes considered. The blue stars are the values of the scaled cross section at the $\jp$ and $\pp$ masses extracted from the 2024~\cite{PhysRevD.110.030001} decay rates $\Gamma\lt \jp\to\LS\rt$ and $\Gamma\lt\pp\to\LS\rt$ using the expressions of Eq.~\eqref{eq:rate-xs} with $\lab A^{\jp,\pp}_{\lsz}\rab=\lab A_{\lsz}(M_{\jp,\pp}^2)\rab$, i.e.,
\be
\tilde\sigma^{\jp,\pp}_\star=\frac{4\pi\alpha^2}{3M_{\jp,\pp}^2}\frac{\Gamma^{\jp,\pp}_{\lsz}}{\Gamma^{\jp,\pp}_{\mu\mu}}\,,
\label{eq:xs-star}
\en
while the red star is the scaled cross section at the $\pp$ mass obtained by the 2018~\cite{PhysRevD.98.030001} decay rate $\Gamma(\pp\to\LS)$. The orange band is the fit function of Eq.~\eqref{eq:fit} with the parameter value given in Eq.~\eqref{eq:A-best}.
\\
The predictions for the scaled cross section at the \jp\ and \pp\ masses are
\be
\begin{array}{rcl}
\tilde\sigma_{\rm fit}(M_{\jp}^2)\!\!\!\!&=&\!\!\!\!\lt 5.79\pm 0.30\rt\,{\rm pb}\,,\\
\tilde\sigma_{\rm fit}(M_{\pp}^2)\!\!\!\!&=&\!\!\!\!\lt 0.825\pm 0.043\rt\,{\rm pb}\,.\\
\end{array}
\label{eq:fit-2024}
\en
On the other hand, using the branching fractions~\cite{PhysRevD.110.030001}
\be
\begin{array}{rcl}
{\rm BR}(\jp\to\LS)\!\!\!\!&=&\!\!\!\! \lt2.83\pm 0.23\rt\times 10^{-5}\,,\\
{\rm BR}(\pp\to\LS) \!\!\!\!&=&\!\!\!\! \lt 1.6\pm 0.7	\rt\times 10^{-6}\,,\\
\end{array}\label{eq:br-2024}
\en
in Eq.~\eqref{eq:xs-star} the corresponding scaled cross sections, represented by blue stars in Fig.~\ref{fig:scaled-xs}, are
\be
\begin{array}{rcl}
\tilde\sigma_{\star}^{\jp}\!\!\!\!&=&\!\!\!\!\lt 6.45\pm 0.53\rt\,{\rm pb}\,,
\\
\tilde\sigma_{\star}^{\pp} 
\!\!\!\!&=&\!\!\!\!
\lt 1.64\pm 0.7	4\rt\,{\rm pb}\,.\\
\end{array}
\label{eq:2024}
\en
It is interesting to notice how much the last value, which refers to the $\pp$ decay, has changed since 2018 when the branching fraction was~\cite{Ferroli:2020xnv}
\be
\!\!\!\!\!\!\!\!{\rm BR}_{18}(\psi(2S)\to\LS)=(1.23\pm0.24)\times 10^{-5}\,,
\label{eq:br-2018}
\en
corresponding to the scaled cross section 
\be
\tilde\sigma_{18\star}^{\pp} &=&\lt 12.6\pm 2.6\rt\,{\rm pb}\,,
\label{eq:2018}
\en
indicated by a red star in Fig.~\ref{fig:scaled-xs}.
\section{Why isospin does strike back}
In light of the drastic change of the scaled cross section, i.e., of the branching fraction BR$(\pp\to\LS)$ which, from 2018, Eq.~\eqref{eq:br-2018}, to 2024, Eq.~\eqref{eq:br-2024}, decreased by a factor of more than seven, two questions arise.
\\
The first concerns the reason of this change and the reliability of the present experimental value of BR$(\pp\to\LS)$.
\\
The second deals with the interpretation of the now reduced discrepancy between the two values of the scaled cross section, Eqs.~\eqref{eq:fit-2024} and~\eqref{eq:2024}, in terms of the effect of a would-be isospin violation amplitude contributing only to the decay. 
\subsection{What has changed since 2018}
\label{subsec:changing}
Undoubtedly, the branching fraction's value for the decay $\pp\to\LS$ published by the 2018 Review of Particle Physics~\cite{PhysRevD.98.030001} is high. The reference value supporting such a quantitative statement is the branching fraction of the decay of the \jp\ into the same final state. This branching fraction reported in Eq.~\eqref{eq:br-2024}, unchanged since 2018, is of the same order, a little more than twice the one of the \pp.\\ 
Using Eq.~\eqref{eq:xs-star}, hence assuming a dominant electromagnetic contribution and the power scaling predicted by perturbative QCD, the expected ratio between $\jp$ and $\pp$ decay rates divided by their electric widths should be of the order of the eight power of the inverse mass ratio, i.e.,
\be
\frac{\Gamma^{\jp}_{\lsz}/\Gamma^{\jp}_{\lsz}}
{\Gamma^{\pp}_{\mu\mu}/\Gamma^{\pp}_{\mu\mu}}
\simeq
\lt\frac{M_{\pp}}{M_{\jp}}\rt^{8}
\simeq 4.0\,.
\label{eq:R-QCD}\en
The 2018 experimental values, considering the 2018 and 2024 data for the width $\Gamma^{\pp}_{\lsz}$ are
\be
\frac{\Gamma^{\jp}_{\lsz}/\Gamma^{\jp}_{\mu\mu}}
{\Gamma^{\pp}_{\lsz}/\Gamma^{\pp}_{\mu\mu}}
=\lgr\begin{array}{lcl} 
0.31\pm 0.07 && 2018~\mbox{\cite{PhysRevD.98.030001}}\\
2.4\pm 1.0 && 2024~\mbox{\cite{PhysRevD.110.030001}}\\
\end{array}\right..
\ne
The 2018 value is not only incompatible with the expectation but goes in the opposite direction being less than one. Instead, the 2024 ratio is greater than one and in fair agreement with the exception of Eq.~\eqref{eq:R-QCD}.
\\
Investigating the reasons behind the drastic difference between the 2018~\cite{PhysRevD.98.030001} and the 2024~\cite{PhysRevD.110.030001} values of the branching fraction $\psi(2S)\to\LS$ is not the aim of this work. However, we notice that in the 2018 edition of the Review of Particle Physics~\cite{PhysRevD.98.030001}, the datum
\be
{\rm BR}_{18}(\psi(2S)\to\LS)=(1.23\pm0.24)\times 10^{-5}\,,
\ne
has a footnote where the authors declare that it had been obtained "using CLEO-c data~\cite{Dobbs:2017hyd} but not authored by the CLEO Collaboration".
\\
Instead, the actual value published in the 2024 edition of the Review of Particle Physics~\cite{PhysRevD.110.030001}, i.e.,
\be
{\rm BR}(\psi(2S)\to\LS)=(1.6\pm0.7)\times 10^{-6}\,
\ne
is the result of the unique analysis by the BESIII experiment~\cite{BESIII:2021mus}, hence it does not consider the CLEO-c data any more.
\subsection{Reinterpretation}
\label{subsec:reinterpretation}
\begin{figure}[t]
	\begin{center}
		\includegraphics[width = .8\columnwidth]{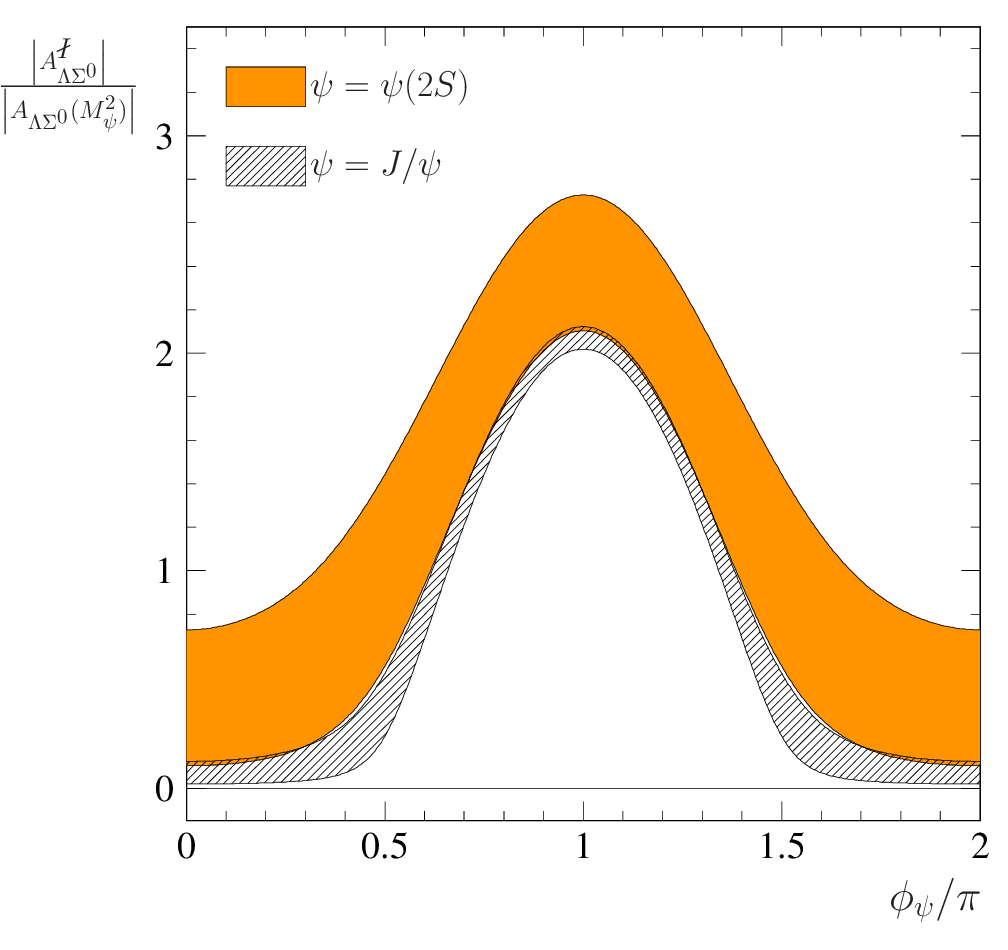}\vspace{-5mm}
		\label{fig:iso-iso}
		\caption{Relative contributions of the isospin-violating amplitude with respect to the electromagnetic one as a function of the relative phase $\phi_\psi$ in units of $\pi$ radiants in the cases of \pp, orange band, and \jp, lined band.}
	\end{center}
\end{figure}

The conclusions of Ref.~\cite{Ferroli:2020xnv} must be revised in light of the latest $\pp\to\LS$ branching fraction value. We had been proposing two hypotheses to explain the high value of the $\pp\to\LS$ branching fraction, giving more weight to that of an uncommon phenomenon of isospin violation guided by the exciting theoretical perspective on one side and by anthropic confidence on the correctness of any human result on the other.  
\\
However, even though the second hypothesis, which we defined as less probable, appears as the right one, our anthropic faith remains unchanged because another and stronger effect emerges overbearingly: isospin symmetry. The lesson we should draw is that a well-established theory is the best yardstick for judging an uncommon result, especially if supported by weak foundations.
\\
With the results of Eq.~\eqref{eq:fit-2024}, the squared values of the ratios defined in Eq.~\eqref{eq:R} for \jp\ and \pp\ are
\be
\lt R_{\lsz}^{\psi}\rt^2
&=& \frac{4\pi\alpha^2}{3M_{\psi}^2\Gamma^{\psi}_{\mu\mu}}\frac{\Gamma^{\psi}_{\lsz}}{\beta_{\lsz}(M_\psi^2)\tilde\sigma_{\rm fit}(M_{\psi}^2)}
\no\\
&=&\lgr
\begin{array}{lcl}
1.15\pm0.11&&\psi=\jp	\\
2.00\pm0.88&&\psi=\pp	\\
\end{array}
\right.,
\label{eq:R-value}\en
where we have used the definition of the complete cross section 
$\sigma_{\lsz}(M_\psi^2)=\beta_{\lsz}(M_\psi^2)\tilde\sigma_{\rm fit}(M_\psi^2)$.
The relative isospin-violating-to-pure-electromagnetic intensities for the \jp\ and \pp\ as functions of the relative phase $\phi_\psi$, as defined in Eq.~\eqref{eq:relativo}, are shown in Fig.~\ref{fig:iso-iso}. The behaviors obtained follow the general trend of Fig.~\ref{fig:example}, which is symmetric with respect to $\phi_\psi=\pi$, where they reach the maximum value equal to $R^\psi_{\lsz}+1$, while the minimum at $\phi_\psi=0$ and symmetrically at $\phi_\psi=2\pi$ radians is 
	$R^\psi_{\lsz}-1$. The minimum is attained when the amplitudes $A^{\noi}_{\lsz}$ and $A_{\lsz}(M_\psi^2)$ are real to each other and have the same sign; instead, the maximum when the amplitudes are still real to each other but have opposite signs. The possibility of significative isospin-violating effects for the \pp, which was the case of study of Ref.~\cite{Ferroli:2020xnv} is drastically reduced considering the new value of the $\pp\to\LS$ branching fraction. The entity of such an effect can be evaluated by the distance of the ratio $R^{\pp}_{\lsz}$ from unity, i.e., by the compatibility with zero of the minima of the curve of Fig.~\ref{fig:iso-iso}. 
	Using the datum of Eq.~\eqref{eq:R-value} for the $\pp$ case the minimum is
	\be
	\left.
\frac{\lab A_{\lsz}^{\noi}\rab}{\lab A_{\lsz}\big(M_{\pp}^2\big)\rab}\right|_{\phi_{\pp}=0}=
 R_{\lsz}^{\pp}-1=0.41\pm0.31\,,
 \ne
 which is compatible with zero within less than two standard deviations.
 The situation for the \jp\ is unchanged since the previous study~\cite{Ferroli:2020xnv}, in this case the minimum is
\be
	\left.
\frac{\lab A_{\lsz}^{\noi}\rab}{\lab A_{\lsz}\big(M_{\jp}^2\big)\rab}\right|_{\phi_{\jp}=0}=
 R_{\lsz}^{\jp}-1=0.07\pm0.06\,,
 \ne
definitely compatible with zero.
\\
In both cases, the possibility of few \% isospin-violating effects remains, even though, as is theoretically expected. However, with the recent measurement of the BESIII Experiment~\cite{BESIII:2021mus} of the $\pp\to\LS$ branching fraction, spectacular isospin violation phenomena are excluded.  Isospin has struck back.






\bibliography{FR-SP-ISB}

\end{document}